\documentclass{jpp}

\usepackage{graphicx}
\usepackage{verbatim}
\usepackage{amsmath}
\usepackage{amssymb}
\usepackage{xcolor}
\usepackage[loading]{tracefnt}
\newcommand{\pd}[2]{\frac{\partial #1}{\partial #2}}
\newcommand{\gyav}[1]{\left\langle #1 \right\rangle_\mathbf{R} }
\newcommand{\vdrift}[2][s]{\ensuremath{\mathbf{v_{d}}_{s}}}

\newcommand{\intd}[1]{ \mathrm{d} #1 \,}

\newcommand{\edit}[1]{#1}
\newcommand{\editb}[1]{#1}
\newcommand{\editc}[1]{#1}

\newcommand{\RN}[1]{%
  \textup{\uppercase\expandafter{\romannumeral#1}}%
}
\usepackage{natbib}

\begin{document}

\title[Transport-modified slowing-down distribution]{\editc{Analytic slowing-down distributions as modified by turbulent transport}}
\author[G. J. Wilkie]{G. J. Wilkie\thanks{Email address for correspondence: wilkie@chalmers.se} }
\affiliation{Department of Physics, Chalmers University of Technology. Gothenburg, Sweden}

%\ead{gwilkie@umd.edu}

%\affiliation{$^1$ Institute for Research in Electronics and Applied Physics, University of Maryland, College Park, MD 20742, USA\\ [\affilskip]
%$^2$ Princeton Center for Theoretical Science, Princeton University, Princeton, NJ 08544, USA \\ [\affilskip]
%$^3$ Rudolph Peierls Centre for Theoretical Physics, University of Oxford, Oxford OX1 3NP, UK }
%\affiliation{Chalmers University of Technology, Gothenburg, Sweden}

\maketitle

\begin{abstract}
   The effect of electrostatic microturbulence on fast particles rapidly decreases at high energy, but can be significant at moderate energy. Previous studies found that, in addition to changes in the energetic particle density, this results in nontrivial changes to the equilibrium velocity distribution. These effects have implications for plasma heating and the stability of Alfv\'en eigenmodes, but make multiscale simulations much more difficult without further approximations. \editc{Here, several related analytic model distribution functions are derived from first principles with reasonable approximations}. A single dimensionless parameter characterizes the relative strength of turbulence relative to collisions, and this parameter appears as an exponent in the model distribution functions. \editc{Even the most simple of these models} reproduces key features of the numerical phase-space transport solution and provides a useful \emph{a priori} heuristic for determining how strong the effect of turbulence is on the redistribution of energetic particles in toroidal plasmas.
\end{abstract}

\section{Introduction}

Energetic particles, such as those utilized to heat plasma in magnetic confinement experiments, are subject to being redistributed by turbulent transport. The net flux of these non-Maxwellian energetic particles between flux surfaces is a strong function of energy \citep{hauff_electrostatic_2009}, and this causes the lower-energy (though still not thermalized) part of the velocity distribution to be ``carved out'', sometimes leaving a bump-on-tail feature \citep{wilkie_transport_2016}. Here, a simplified model which describes this effect is introduced, with a particular focus on alpha particles produced by the deuterium-tritium (DT) fusion reaction. It is a generalization of the analytic slowing-down distribution of Gaffey \citep{gaffey_energetic_1976,helander_collisional_2002} (which will repeatedly be referred to as the ``classical'' slowing-down distribution) that includes an additional term mimicking the velocity dependence of microturbulent transport. \edit{An analytic distribution was also previously derived by \cite{anderson_influence_1991} and \cite{sigmar_effects_1993}, wherein anomalous transport was treated with a constant diffusion coefficient, rather than one with velocity dependence.}

Alpha particles are born isotropically from reacting deuterium and tritium nuclei at an energy of $E_\alpha = m_\alpha v_\alpha^2 / 2 = 3.52 \, \mathrm{MeV}$. Plasma heating is caused by the friction of these particles against the bulk ions/electrons  (which have characteristic thermal speeds $v_{ti/e} = \sqrt{2 T_{i/e} / m_{i/e}}$, respectively). As they slow down and their magnetic orbits become smaller, alpha particles become more and more susceptible to ion-scale microturbulence. As they transport outward, they continue slowing down on outer flux surfaces. This, combined with the fact that fewer alpha particles are produced near the cooler edge of the plasma, means that the alpha particle energy leaving the plasma will be relatively be small \citep{kurki-suonio_fast_2011}. Even so, alpha particles will be redistributed radially, changing the global heating profile. Alphas then eventually become thermalized in equilibrium with the bulk ions, creating a population of helium ``ash'' with density $n_\mathrm{ash} > n_\alpha$\footnote{Here, $n_\alpha$ refers to the density of helium that is \emph{not} Maxwellian ash so that the total helium density is $n_\mathrm{He} = n_\mathrm{ash} + n_\alpha$} and temperature $T_\mathrm{ash} = T_i$. This population can then be treated as a Maxwellian impurity.

All of the aforementioned effects are captured by the kinetic transport equation from the low-collisionality ordering of gyrokinetics for the equilibrium distribution $f_\alpha$:
\begin{align} \label{transporteqn}
   \pd{f_\alpha}{t} -& \frac{1}{V'} \pd{ }{r} \left( V' D_{rr} \pd{f_\alpha}{r} + V' D_{rv} \pd{f_\alpha}{v} \right) - \frac{1}{v^2} \pd{ }{v} \left( v^2 D_{vr} \pd{f_\alpha}{r} + v^2 D_{vv} \pd{f_\alpha}{v} \right)  \nonumber \\
    &=  C \left[ f_\alpha \right] + S_\alpha.
\end{align}
The collision operator $C$ includes collisions of trace particles against the Maxwellian bulk, and $S_\alpha(r,v)$ is the source of alpha particles from fusion reactions. Flux surfaces are labelled by the radial coordinate $r$, defined as the surface half-width at the height of the magnetic axis, and enclose a volume $V(r)$, and primes denote differentiation with respect to radius. The radius of the last closed flux surface is the plasma minor radius $a$. The various turbulent diffusion coefficients $D_{xx}$ are functions of radius and speed $v$, and these can be calculated from nonlinear gyrokinetic simulations. Equation \eqref{transporteqn} is valid  for either trace impurities \emph{or} for energetic particles \citep{wilkie_2018}. Both of these conditions are typically satisfied for alpha particles \citep{wilkie_validating_2015}. 
%In this work, numerical solutions to equation \eqref{transporteqn} are calculated with \textsc{t3core}, a phase-space transport solver \citep{wilkie_transport_2016}.

%\editc{The main body of this work derives (Section \ref{section2}), benchmarks, and discusses (Section \ref{section3}) a particularly simple analytic distribution using a local loss term with the approximate velocity dependence of isotropic turbulent transport. The approximate velocity dependence used is derived from first principles in Appendix \ref{appA}. Appendix \ref{appB} relaxing the isotropic approximation and leads to a velocity dependence with the same energy scaling, but a different pitch-angle scaling than \ref{hauff_electrostatic_2009}. An analytic model distribution with pitch angle dependence is thereby also derived in Appendix \ref{appB}. Appendix C relaxes the local-sink approximation to solve analytically for the radial transport in a plasma that is uniform except for the energetic particle source.}
\editc{The main body of this work derives (Section \ref{section2}), benchmarks, and discusses (Section \ref{section3}) a particularly simple analytic distribution using a local loss term that models the approximate effect of energy-dependent transport. The approximate velocity dependence used is derived from first principles in Appendix \ref{appA}. Appendix \ref{appB} relaxes the isotropic approximation and leads to a velocity dependence with the same energy scaling, but a different pitch-angle scaling than \cite{hauff_electrostatic_2009}. An analytic model distribution with pitch angle dependence is thereby also derived in Appendix \ref{appB}. Appendix C relaxes the local-sink approximation to solve analytically for the radial transport in a plasma that is uniform except for the energetic particle source. These generalizations are combined to obtain an analytic distribution in 3-dimensional phase space $f_{SD,3D}\left(r, v, \xi\right)$ in Eq. \eqref{fullsoln}.}

\section{Model distribution function} \label{section2}

An approximate local solution to Eq. \eqref{transporteqn} will be found after a series of assumptions. 
%First, simplify the source to be mono-energetic: $S_\alpha \approx \left(S_0/4\pi v_\alpha^2 \right) \delta\left(v - v_\alpha \right)$. \edit{In actuality, there will be a significant spread in energy, even for fusion produced alpha particles \citep{brysk_fusion_1973}, but a delta source is used here to simplify the analysis. 
\edit{In this section, it will be assumed that the energetic particles are isotropic in velocity space, but this is relaxed in the more general treatment of Appendix \ref{appB}.} Also, in the limit where $v_{ti} \ll v \ll v_{te}$, the collision operator acting on an isotropic alpha distribution reduces to the simplified form $C\left[ f_\alpha \right] \approx \tau_s^{-1} v^{-2} \left(\partial/ \partial v \right) \left( v_c^3 + v^3 \right) f_\alpha$, where the critical speed is defined as
\begin{equation} \label{vcdef}
   v_c \equiv v_{te} \left( \frac{3 \sqrt{\pi}}{4} \sum\limits_i \frac{n_i m_e}{n_e m_i} Z_i^2 \right)^{1/3},
\end{equation}
with the sum over all non-trace ionic species. The slowing-down time is: 
\begin{equation} \label{tausdef}
   \tau_s \equiv \frac{3}{16 \sqrt{\pi}} \frac{m_\alpha m_e v_{te}^3}{Z_\alpha^2 e^4 n_e \ln\Lambda_{\alpha e}}.
\end{equation}

Previous studies \citep{wilkie_global_2017} found that the $D_{rv}$, $D_{vr}$ and $D_{vv}$ coefficients play a sub-dominant role compared to radial diffusion ($D_{rr}$) at high energy, so this will be the only transport term kept from Eq. \eqref{transporteqn}. To see that this is justified, consider the two source terms in the gyrokinetic equation: $\omega e \gyav{\phi} \partial f_\alpha / \partial E$ and $(c/B) k_y \gyav{\phi} \partial f_\alpha / \partial r$ for the case of ion-scale turbulence ($\omega \sim v_{ti}/a$ is the frequency, $k_y \sim \rho_i^{-1}$ is a perpendicular wave number, $\rho_i$ is the thermal ion Larmor radius, $E$ is energy, $B$ is the magnetic field magnitude, and $\gyav{\cdot}$ denotes the gyroaverage). The former term contributes to $D_{rv}$ and the latter to $D_{rr}$. Therefore, the latter will dominate when $T_i \partial f_\alpha / \partial E \ll a \partial f_\alpha / \partial r$, as is the case for energetic particles. Similar arguments apply to $D_{vr}$ and $D_{vv}$ when comparing their definitions to $D_{rr}$ and $D_{rv}$ \citep{wilkie_transport_2016}. \edit{See Appendix \ref{appA} for more details.}

\editc{An analytic model is derived in Appendix \ref{globalapp} which includes the global radial diffusion operator, but includes other generous assumptions about the radial plasma profile. Here,} 
 a \emph{local} approximation will be made for the radial diffusion term, where the radial derivatives are parametrized by a length scale such that:
\begin{equation} \label{localapprox}
   \frac{1}{V'} \pd{ }{r} \left[ V' D_{rr} \pd{f_\alpha}{r} \right] = -\frac{D}{L_\alpha^2} f_\alpha.
\end{equation}
This is done to capture the approximate effect of transport on the local velocity distribution without needing to solve the radial transport equation as was done in \cite{wilkie_transport_2016}. In this framework, transport acts as a velocity-dependent \emph{sink} of alpha particles. In reality, transported particles move to other flux surfaces, where they continue to slow down. Therefore, this approximation represents a ``worst-case scenario''. 
%As severe as this approximation is, it results in a useful parametrization for the effect of energy-dependent transport on the slowing-down distribution. 
This excludes cases where the energetic particle density might be \emph{greater} at some radii due to transport from inner flux surfaces. \edit{In such cases, a local model is inadequate and full radial transport solution of Eq. \eqref{transporteqn} is required \citep{wilkie_transport_2016}.}

Since one does not have the luxury of the full radial and velocity dependence of $f_\alpha$, an \emph{a priori} proxy is desired for the purposes of defining $L_\alpha$. For the classical slowing-down distribution, $n_\alpha \propto S_0$, so the source is chosen to represent the radial derivative of the energetic particle distribution so that $\partial f_\alpha / \partial r \approx -f_\alpha /L_\alpha$. This defines the radial scale length:
\begin{equation} \label{Ldef}
   L_\alpha^{-2} \equiv -\mathrm{max} \left[ \frac{S_0''}{S_0} + \frac{S_0'}{S_0} \left( \frac{D_\alpha'}{D_\alpha} + \frac{V'}{V} \right), 0 \right],
\end{equation}
where $D_\alpha = D_{rr}\left( v= v_\alpha \right)$ is typically small, less than about $0.01 \mathrm{m}^2/\mathrm{s}$. \editc{This form of $L_\alpha$ is kept in the main body of this work for generality, but it does have the drawback of choosing a relevant energy to represent the radial distribution of energetic particles. For an analytic model that relaxes this assumption, see Appendix \ref{globalapp}.}
\edit{ The neoclassical diffusion of slowing-down alphas due to collisional pitch-angle scattering \citep{catto_evaluation_1987} is expected to even smaller than turbulent transport, though this may be enhanced by a lack of perfect toroidal symmetry near the plasma edge \citep{kurki-suonio_fast_2011}. }
   
\edit{The model energy dependence of the diffusion coefficient used for this model is borrowed from \cite{hauff_electrostatic_2009}:
   \begin{equation} \label{Dmodeldef}
   D_{rr} = D_\alpha \frac{v_\alpha^3}{v^3}.
   \end{equation}
See Appendix \ref{appA} for details on how this scaling is found. In Appendix \ref{appB}, it is shown that this is good approximation to the more general case where the pitch-angle dependence of $D_{rr}$ is taken into account. This is because the range of pitch angles for which $D_{rr}$ departs from this $v^{-3}$ scaling is very narrow \citep{pueschel_anomalous_2012}.} 

One benefit of the locality approximation is that it reduces the effect of transport to a single dimensionless parameter, which shall be defined as:
\begin{equation} \label{bdef}
   b \equiv  \frac{D_\alpha \tau_s}{L_\alpha^2} \frac{v_\alpha^3}{v_c^3}.
\end{equation}
The appearance of the factor $v_\alpha^3/v_c^3$ is curious, but is motivated by the solution that follows. 
This dimensionless parameter can be simplified further by scaling the diffusion coefficient similarly to \cite{pueschel_anomalous_2012}. There, the diffusion coefficient at thermal energies was deemed proportional to the effective thermal diffusivity $\chi_\mathrm{eff}$. If it is supposed that $D_{rr}\left(v=v_c\right) \approx \chi_\mathrm{eff}$, then one can approximate $b \approx \chi_\mathrm{eff} \tau_s / L_\alpha^2 $. 
%More importantly, it was found that at low $\beta$, the diffusion coefficient scales like $v^{-3}$ for non-extreme pitch angles (see Appendix \ref{appA} for more details). Therefore, $D_{rr}\left(v \right) = D_\alpha v_\alpha^3 / v^3$ is used in the model that follows.

With the aforementioned approximations, Eq. \eqref{transporteqn} reduces to:
\begin{equation} \label{approxtransporteqn}
%    \frac{b}{\tau_s} \frac{v_c^3}{v^3} f_\alpha = \frac{1}{\tau_s} \frac{1}{v^2} \pd{ }{v} \left[ \left( v_c^3 + v^3 \right) f_\alpha \right] + \frac{S_0}{4 \pi v_\alpha^2} \delta\left( v - v_\alpha \right)
   \frac{b}{\tau_s} \frac{v_c^3}{v^3} f_\alpha = \frac{1}{\tau_s} \frac{1}{v^2} \pd{ }{v} \left[ \left( v_c^3 + v^3 \right) f_\alpha \right] + \edit{S_\alpha}
\end{equation} 
in steady-state. In this model, the transport term diverges as $v \rightarrow 0$. More realistically, one could define a cutoff speed below which the diffusion coefficient is constant -- this also results in an analytic solution.
%In that case, the solution to Eq. \eqref{approxtransporteqn} is:
%\begin{equation} \label{Fsdmodef}
%   F^\mathrm{(cutoff)}_{SD,\mathrm{mod}} = \ldots.
%\end{equation}
However, due to the other approximations made, Eq. \eqref{approxtransporteqn} is invalid at low energy, where the distribution is dominated by thermalized helium ash. Therefore, a bare $D_{rr} \propto v^{-3}$ scaling, without an ad-hoc cutoff, is used without significant reservation. This is done with the knowledge that, like the classical slowing-down distribution, it is not valid for $v \lesssim v_{ti}$. Unfortunately, this precludes a useful analogous generalization for heated minority ions \citep{pusztai_turbulent_2016} since it is much more sensitive to the diffusion coefficient at low energy, for which a reduced model has yet to be developed. However, the generalization to injected auxiliary fast ions (e.g. neutral beams) is straightforward, even if their injection energy is not as high as for alpha particles.

\iffalse
\editb{Without resorting to defining a local scale length $L_\alpha$, under certain assumptions about the radial profile, one can instead find a global analytic distribution. Let the source of alpha particles has a radial dependence that vanishes at the plasma edge with a Bessel function dependence: $S_\alpha = S_0 J_0\left( a_{0,1} r / a \right)$, where $S_0$ is now the value on the magnetic axis. The first zero of $J_0$ is $c_{0,1}$. Suppose further that $\tau_s$, $v_c$, $D_\alpha$ are radially homogeneous and that the flux surfaces are circular so $V'(r) = 2 \pi^2 R r$. Now, Eq. \eqref{transporteqn} (with $D_{rv} = D_{vr} = D_{vv} = 0$) can be solved directly be expanding in a Fourier-Bessel series. Then, $f_\alpha$ has the same radial dependence as the source and now the effective transport parameter $b = \tau_s c_{0,1}^2 D_\alpha v_\alpha^3/ v_c^3 a^2$. }
\fi

\begin{figure}
   \begin{center} \includegraphics[width=0.4\textwidth]{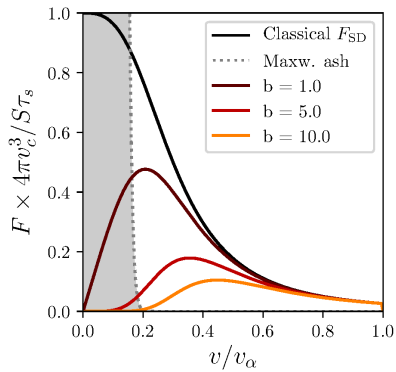} \end{center}
   \caption{\label{Fsdmod} Comparing the classical Gaffey slowing-down distribution with the analytic transport-modified version -- Eq. \eqref{Fsdmodef} -- for different values of $b$. Lighter shades corresponding to higher relative turbulence intensity. The shaded area represents the Maxwellian ash population for $n_\mathrm{ash} = 10 n_\alpha$ and $T_\mathrm{ash} = E_\alpha / 350$. For these cases, $v_c = 0.3 v_\alpha$. }
\end{figure}

The solution to Eq. \eqref{approxtransporteqn} is:
\edit{
\begin{equation} \label{Fsdgensource}
   f_\alpha = \frac{\tau_s}{v_c^3 + v^3} \left( \frac{v^3}{v^3 + v_c^3} \right)^{b/3 } \int_{\infty}^v S_\alpha\left( v' \right) \left( \frac{v'^3}{v'^3 + v_c^3} \right)^{-b/3 } \, \mathrm{d} v'.
\end{equation}
Now the choice of including $v_\alpha^3 / v_c^3$ in the definition of $b$, Eq. \eqref{bdef}, becomes clear: this factor appears in the exponent of this distribution, and is thereby important in quantifying the effect of turbulence on the energetic particle distribution. Even for fusion-produced alpha particles, the source $S_\alpha$ has a significant spread in energy proportional to the geometric mean of the fuel temperature and the alpha particle birth energy \citep{brysk_fusion_1973}. However, in order to obtain an analytic solution, it will be convenient to use a mono-energetic source such that $S_\alpha \approx \left(S_0/4\pi v_\alpha^2 \right) \delta\left(v - v_\alpha \right)$. The following analytic model distribution is thereby found: }
\begin{equation} \label{Fsdmodef}
   F_{SD,\mathrm{mod}} = \frac{S_0 \tau_s}{4 \pi} \frac{1}{v_c^3 +v^3 } \left( \frac{v^3}{v_\alpha^3} \frac{v_\alpha^3 + v_c^3}{v^3 + v_c^3} \right)^{b\edit{/3} }  H\left( v_\alpha - v \right),
\end{equation}
where $H$ is the Heaviside step function. Equation \eqref{Fsdmodef} reduces to the classical slowing-down distribution when transport is negligible: $b = 0$. \edit{This is a generalization of the analytic distribution found by \cite{sigmar_effects_1993} to the case of energy-dependent diffusion as given by Eq. \eqref{Dmodeldef}.} Representative examples of this distribution at different values of $b$ are shown in Fig. \ref{Fsdmod}. 
%Note that even when $D_\alpha \tau_s / L_\alpha^2$ is relatively small, this can have a significant effect on the alpha particle distribution at moderate energy $v\sim v_c$. This is due to the additional factor of $v_\alpha^3 / v_c^3$ in the exponent of Eq. \eqref{Fsdmodef}, which motivated the definition of $b$ in Eq. \eqref{bdef}. 
%Therefore, Eq. \eqref{Fsdmodef} is kept as the model with the knowledge that, like the classical slowing-down distribution, it is not valid for $v \lesssim v_{ti}$.

\section{Benchmarking and discussion} \label{section3}

There are several curious features of the distribution described by Eq. \eqref{Fsdmodef} which are nevertheless backed up by numerical simulations and physical insight. 

Firstly, $F_{SD,\mathrm{mod}}(v)$ has a local maximum at $v_* =  v_\alpha \left( D_\alpha \tau_s / 3 L_\alpha^2 \right)^{1/3}$. Therefore, the distribution exhibits a ``bump-on-tail'' similar to what was previously observed \citep{wilkie_transport_2016,wilkie_global_2017}. In order for this inversion to be observable, this maximum must be well separated from the helium ash population. Therefore, a bump-on-tail caused by turbulent transport is possible only if:
\begin{equation} \label{bumpcondition}
   D_\alpha \gg \frac{3 L_\alpha^2}{\tau_s} \left( \frac{v_{t,\mathrm{ash}}}{v_\alpha} \right)^3.
\end{equation}
Using $T_i \sim 10 \mathrm{keV}$, $\tau_s \sim 1 \mathrm{s}$, and $L_\alpha \sim 0.4 \mathrm{m}$, the presence of a bump requires $D_\alpha \gg 7 \times 10^{-5} \mathrm{m}^2/\mathrm{s}$. \edit{Estimates of $D_\alpha$ from the literature range from about $6 \times 10^{-4}$ to $2 \times 10^{-2} \mathrm{m}^2/\mathrm{s}$ using gyrokinetic simulations \citep{hauff_electrostatic_2009,zhang_scalings_2010,pueschel_anomalous_2012,wilkie_validating_2015}, making this condition marginally- or well-satisfied depending on the scenario. The presence of a local maximum in velocity space has the potential to destabilize waves, such as Alfv\'en eigenmodes, via inverse Landau damping. A flattening or inversion in velocity space is destabilizing, but it accompanied by a flattening of the radial profile, which is stabilizing. Without a detailed analysis, it is not immediately clear what the net effect is.}

%In D-T experiments, the line-of-sight distribution of alpha particles also exhibited a bump-on-tail feature upon measurement, but at rather high energy: $v_* \approx 0.6 v_\alpha$ \citep{sharapov_burning_2008}. In order to ascribe this feature to microturbulence, this requires $b \approx 0.6$. Assuming $L_\alpha \approx 0.15 \mathrm{m}$ and $\tau_s \approx 1 \mathrm{s}$, this results in $D_\alpha \approx 0.015$, which is a reasonable but relatively large value.

%Another interesting property of the model distribution is that, even when $b$ is relatively small, the exponent in Eq. \eqref{Fsdmodef} gets promoted by a factor of $v_\alpha^3 / v_c^3$.

Also, notice that, at the birth speed $v = v_\alpha$, there is no turbulent correction to the classical slowing-down distribution. No matter how strong the turbulence is, it will not affect newly-born alpha particles because it cannot possibly compete with a delta function at its peak. As the alpha particle slows down, it becomes doubly affected because it is further away in velocity space from the source in addition to turbulent transport becoming stronger. Since most of the electron heating occurs near the injection speed, this remains immune to turbulent transport of alphas. What might be affected is the heating of ions at lower energy.

The model distribution was benchmarked against a numerical solution of Eq. \eqref{transporteqn}. The test case is an ITER-like ELMy H-mode DT scenario that is unstable to electrostatic ion-temperature gradient turbulence beyond $r = 0.5 a$. It can be found on the public tokamak profile database \citep{roach_2008_2008} as ITER scenario 10010100. \editb{In these simulations, the diffusion coefficient was calculated directly from first-principles gyrokinetic simulations (and thus did not diverge as $v \rightarrow 0$). Further, the full test-particle collision operator was used to capture physically consistent thermalization of helium ash.} For more details about the transport of alpha particles in this case, see \cite{wilkie_transport_2016}.

\begin{figure}
   \begin{center} \includegraphics[width=0.6\textwidth]{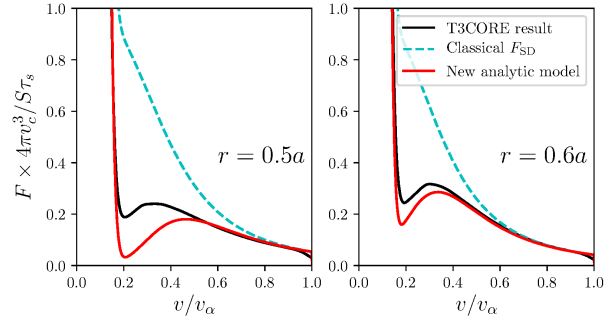} \end{center}
   \caption{\label{iterFsdcomp} Comparing the local velocity distribution function of alpha particles for two different radii of the ITER-like scenario of \citep{wilkie_transport_2016}. Shown are the Gaffey slowing-down distribution (cyan dashed line), Eq. \eqref{Fsdmodef} (red line), and a numerical phase-space transport solution (black line). Each of the analytic distributions are shown with a population of Maxwellian ash added, equal to the ash density calculated with \textsc{t3core}. }
\end{figure}

The local distribution function at two different radii are shown in Fig. \ref{iterFsdcomp}. Despite the generous assumptions built into the analytic model, it qualitatively captures the basic features of the numerical solution including: well-confined high-energy alphas, a depletion of the distribution at more moderate energies, and the bump-on-tail feature. At these radii ($r = 0.5a$ and $0.6a)$, the values of $b$ were found to be 5.3 and 2.6, respectively.

%The density moment of the model distribution differs from that of the Gaffey slowing-down distribution, and. 

\begin{figure}
   \begin{center} \includegraphics[width=0.4\textwidth]{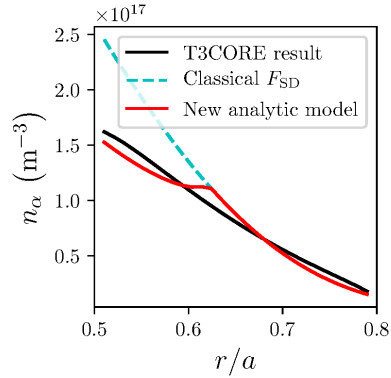} \end{center}
   \caption{\label{densFsd} Comparing the radial density profile of energetic alpha particles as predicted by the Gaffey slowing-down distribution (cyan dashed line), Eq. \eqref{Fsdmodef} (red line), and a numerical phase-space transport solution (black line). }
\end{figure}

The alpha particle density as a function of radius for the test case is shown in Fig. \ref{densFsd}. There, the analytic solution is compared to the numerical transport solution and to the slowing-down distribution without radial transport. 
%The density moment of the model distribution differs from that of the Gaffey slowing-down distribution, and this difference is shown in Fig. \ref{densFsd}. 
Even when the classical density differs from the numerical one due to transport, Eq. \eqref{Fsdmodef} captures this difference fairly well. This motivates using the model as a heuristic to answer the question: ``can one expect significant energetic particle transport from microturbulence?'' An affirmative answer indicates that further study is warranted, but if $b < 1$, the classical slowing-down distribution is reasonably accurate.

\section{Conclusion}

This work introduced and briefly discussed a simplified model for slowing-down alpha particles in the presence of microturbulence. Although heuristic, this model distribution captures the approximate effect of turbulence through a single dimensionless parameter, which has a stronger impact than one might expect when the birth speed $v_\alpha$ significantly exceeds the critical speed $v_c$. \editc{When relaxing several assumptions made for this simple model, more robust analytic distributions are found which maintain the same basic character of Eq. \eqref{Fsdmodef}.} \editc{Although more detailed simulations are generally required to robustly predict the impact of microturbulence on plasma heating, Alfv\'en eigenmode drive, and other effects mediated by energetic particles, this model provides physical insight into \editc{the expected effect of microturbulence on these phenomena}.}

The author would like to thank I. Pusztai, I. Abel, and T. F\"ul\"op for helpful comments in preparing this manuscript. The author was supported by the EUROfusion Researcher Grant AWP18-ERG-VR.  The views and opinions expressed herein do not necessarily reflect those of the European Commission. 

\iffalse
The flattening of the alpha particle distribution in radius and energy has implications for the stability of Alfv\'en eigenmodes. The alpha-particle-induced drive toroidicity-induced Alfv\'en eigenmodes for a simplified case \citep{fu_excitation_1989,ghantous_1.5d_2012} is proportional to:
\begin{equation}
   A_{\alpha,m} = \int \mathrm{d}^3\mathbf{v} \left( v_\|^2 + \frac{v_\perp^2}{2} \right) \delta \left(\omega - k_{\|,m} v_\| \right) \left( \omega E_\alpha \pd{f_\alpha}{E} - \omega_{*,\alpha} a \pd{f_\alpha}{r} \right).
\end{equation}
Calculating the $\partial f_\alpha / \partial r$ term as a function of energy requires a proper treatment of radial transport. But the effect of transport on $\partial f_\alpha / \partial E$ can be estimated with the model distribution \eqref{Fsdmodef}. In Fig. \ref{taedrive} we show the relevant quantity:
\begin{equation}
   B_{\alpha,m} = 2 \pi \int \mathrm{d}v_\perp v_\perp \left( \frac{ \omega^2}{k_{\|,m}^2} + \frac{v_\perp^2}{2} \right) \left( \pd{f_\alpha}{E} \right)_{v_\| = \omega / k_{\|,m}}
\end{equation}
\fi

\appendix

\section{\editb{Energy scaling of the diffusion coefficient}} \label{appA}

In this section, the fluctuating distribution of energetic particles is analyzed to motivate the energy scaling of the diffusion coefficient used in this work \citep{hauff_electrostatic_2009}. Only electrostatic fluctuations will be considered here. After finding a solution to the gyrokinetic equation in the high-energy limit as was done in \cite{wilkie_2018}, the radial flux is averaged over the spatial and temporal scale of the fluctuations. The diffusion coefficient is what remains after factoring out $\partial F_{0\alpha} / \partial r$. In the mean time, it will become apparent why radial diffusion is dominant over the other turbulent terms of Eq. \eqref{transporteqn} in this limit.

The collisionless gyrokinetic equation for non-Maxwellian fast particles is \editb{\citep{frieman_nonlinear_1982,wilkie_microturbulent_2015,abel_unpublished_2018}}:
\begin{equation} \label{gkeqn}
   \pd{h_\alpha}{t} + v_\| \mathbf{b} \cdot \nabla h_\alpha + \left( \mathbf{v}_D  + \mathbf{v}_\phi \right) \cdot \nabla h_\alpha = Z_\alpha e \pd{F_{0\alpha}}{\mathcal{E}} \pd{\gyav{\phi}}{t} - \mathbf{v}_\phi \cdot \nabla F_{0\alpha},
\end{equation}
where it is expressed in the coordinates $\mathcal{E} = Z_\alpha e \phi + m_\alpha v^2 / 2$, and $\mu$, which is the exactly conserved magnetic moment. With this definition, $\partial F_{0\alpha} / \partial \mu$ does not appear in Eq. \eqref{gkeqn}. It does appear, along with the adiabatic contribution, in the equations that determine the fluctuating fields, which will not be used here. The distribution is decomposed into an equilibrium part $F_{0\alpha}$ and a fluctuating part $h_\alpha$ such that $f_\alpha = F_{0\alpha} + h_\alpha$\footnote{The adiabatic contribution to $\delta f_\alpha$, which does not contribute to the flux, comes from Taylor expanding in the small difference between $\mathcal{E}$ and the kinetic energy when taking moments of $f_\alpha$.}. The magnetic field is given in the Clebsch representation as $\mathbf{B} = B \mathbf{b} =  \nabla \alpha \times \nabla \psi$, where $\psi$ is the poloidal flux in the flux surface that it labels and $\alpha$ (not to be confused with the subscript denoting energetic particles) labels the magnetic field line within said flux surface. The velocity parallel to the magnetic field is $v_\| = \sigma_\| \sqrt{\left(2 / m_\alpha \right)\left( \mathcal{E} - \mu B\left( \theta \right) \right) }$ and $\theta$ is the ballooning-extended poloidal angle, used as a coordinate along the field line. 
The fluctuating static potential is $\phi$. The gyroaverage at constant gyrocenter $\mathbf{R}$ is denoted $\gyav{ }$, and the spatial gradients in Eq. \eqref{gkeqn} are taken with respect to $\mathbf{R}$. The drift associated with the fluctuating fields is given by \editb{$\mathbf{v}_\phi \equiv \left(c / B\right) \mathbf{b}\times \nabla \gyav{\phi}$}, and $\mathbf{v}_D$ is the magnetic drift due to the curvature and gradient in $\mathbf{B}$.

Consider that, in ion-scale turbulence, the bulk ions set the spatial and temporal scales for the variation of $\phi$ and $h_\alpha$. In the limit where $v_{t\alpha} \gg v_{ti}$, the magnetic drift, parallel streaming, and equilibrium gradient are the largest terms in Eq. \eqref{gkeqn}. The ``nonlinear'' $\mathbf{v}_\phi$ term is thereby small. It is convenient and conventional to express the fluctuating quantities in an eikonal representation so that $\phi = \hat{\phi} \left( \theta \right) \exp\left[ i S\left(\psi,\alpha \right) \right]$, where $S$ contains the small-scale variation perpendicular to the magnetic field and $\hat{\phi}$ is the relatively long-scale parallel variation \citep{antonsen_kinetic_1980}. In this representation, the energetic limit of Eq. \eqref{gkeqn} reduces to:
\begin{equation} \label{approxgk}
   v_\| \mathbf{b} \cdot \nabla \theta \pd{\hat{h}_\alpha}{\theta} + i \mathbf{v}_D \cdot \nabla S \hat{h}_\alpha = ic \pd{S}{\alpha} \pd{F_{0\alpha}}{\psi} \hat{\phi},
\end{equation}
Even near the source, where $\partial F_{0\alpha} / \partial \mathcal{E} \sim F_{0\alpha}/ T_i$, the first term on the right hand side of Eq. \eqref{gkeqn} (which ultimately contributes to $D_{rv}$) is smaller than the first term (which leads to $D_{rr}$) on the right hand side of Eq. \eqref{approxgk} by virtue of the large spatial gradients  in the source of energetic particles (and therefore in $F_{0\alpha}$ itself). In any case, at energies this high, collisions acting on the equilibrium dominate both $D_{rr}$ and $D_{rv}$. Away from the source, where turbulence is observed to have an effect, $\partial F_{0\alpha} / \partial \mathcal{E} \sim F_{0\alpha}/E_\alpha$, and this term is smaller still. Similar arguments apply to the turbulent heating terms ($D_{vr}$ and $D_{vv}$), only moreso. Furthermore, if the $\partial \gyav{\phi}/ \partial t$ term were kept, its contribution to the time-averaged radial flux vanishes in the energetic (thereby, linear) limit because $\int \phi \dot{\phi} \,\mathrm{d} t = 0$ in steady-state by definition. It is for these reasons that radial diffusion is dominant over other terms in Eq. \eqref{transporteqn}.  

\editb{In finding an integrating factor to solve Eq. \eqref{approxgk}, it will be convenient to define: } 
\begin{equation} \label{zdef}
   z\left( \theta \right) \equiv \int^\theta_{\theta_0} \frac{\omega_D' \,\mathrm{d}\theta'}{v_\parallel' \left( \mathbf{b} \cdot \nabla \theta \right)'},
\end{equation}
where $\omega_D \equiv \mathbf{v}_D \cdot \nabla S$ and $\omega_D\left( \theta_0 \right) = 0$. Primes denote quantities taken at $\theta'$ rather than $\theta$. A saddle point is located at $\theta_0$ such that  $\left[ \mathrm{d} z / \mathrm{d} \theta \right]_{\theta_0} = 0$. 
With this definition, the solution to Eq. \eqref{approxgk} is \citep{kim_electromagnetic_1993,wilkie_2018}:
\begin{equation} \label{hsoln}
   \hat{h}_\alpha\left( \theta \right) = i c  \pd{F_{0\alpha}}{\psi} \pd{S}{\alpha} \int^\theta J_{0}\left( \frac{| \nabla S | v_\perp}{\Omega_\alpha \left( \theta' \right) } \right)  \phi \left( \theta' \right) \exp\left[  iz\left(\theta \right) - i z\left( \theta' \right) \right] \frac{\intd{\theta'}}{v_\parallel'\left( \mathbf{b} \cdot \nabla \theta \right)'},
\end{equation}
where $J_0$ is the zeroth-order Bessel function of the first kind and $\Omega_\alpha \equiv Z_\alpha e B / m_\alpha c$ is the gyrofrequency. \editb{The lower limit of integration in Eq. \eqref{hsoln} is either $\pm \infty$ for passing particles depending on the sign of $v_\|$, or $\pm \theta_b$ if the particles are trapped. Expanding $z$ in a Taylor series about $\theta_0$} leaves a Gaussian which is integrated over the contour of steepest descent, taking advantage of the fact that $z \gg 1$ to obtain: 
\begin{equation} \label{happrox}
   \hat{h}_\alpha\left( \theta \right) \approx i c  \pd{F_{0\alpha}}{\psi} \pd{S}{\alpha} e^{i \left( z_0 - z \right)} J_0\left( \frac{| \nabla S | v_\perp}{\Omega_\alpha \left( \theta_0 \right)} \right)   \frac{\phi\left( \theta_0 \right) }{\left[ v_\parallel \left( \mathbf{b} \cdot \nabla \theta \right) \right]_0} \sqrt{ \frac{2 \pi i}{\zeta_0} },
\end{equation}
where $\left[ \, \right]_0$ denotes quantities taken at $\theta_0$, and $\zeta_0$ is defined as:
\begin{equation} \label{zetadef}
   \zeta_0 \equiv \left[ \frac{ \mathrm{d}^2 z}{\mathrm{d}\theta^2 }\right]_{\theta_0} = \left[ \frac{ \mathrm{d}}{\mathrm{d}\theta} \frac{ \omega_D}{v_\| \mathbf{b} \cdot \nabla \theta } \right]_{\theta_0} .
\end{equation}
\editb{The approximate solution of Eq. \eqref{happrox} excludes deeply trapped particles; namely those that bounce at a poloidal location $|\theta_b| \lesssim |\theta_0|$}.

The radial flux \editb{is kept as a function of velocity in the kinetic phase space transport equation}, and is defined as:
\begin{equation} \label{fluxdefh}
   \Gamma_\alpha \equiv \left\langle \sum\limits_{\sigma_\|} \int_{-\infty}^{\infty} h_\alpha \mathbf{v}_\phi \cdot \nabla r \, \mathrm{d} \theta \right\rangle_{\perp,t},
\end{equation}
where $\left \langle \, \right\rangle_\perp$ is a perpendicular spatial average over the flux tube and a time average over the the fluctuation timescale \citep{abel_multiscale_2013}. In performing the $\theta$ integral, apply the same large-$z$ approximation that led to Eq. \eqref{happrox}. Also note that $h_\alpha$ is proportional to $\partial F_{0\alpha} / \partial \psi$, allowing one to write $\Gamma_\alpha = - D_{rr} \partial F_{0\alpha} / \partial r$ for $\psi = \psi\left(r\right)$.\footnote{This is a specific instance of a more general result that allows Eq. \eqref{transporteqn} to be written as such.} This gives:
\begin{equation} \label{fluxapprox}
   D_{rr} \approx \left\langle 2 \pi c^2  \left(\frac{\mathrm{d} r}{\mathrm{d} \psi}\right)^2 \left(\pd{S}{\alpha} \right)^2  J_0^2\left( \frac{| \nabla S | v_\perp}{\Omega_\alpha \left( \theta_0 \right)} \right)   \frac{ \left| \phi\left( \theta_0 \right) \right|^2 }{ \left| \left( v_\| \mathbf{b} \cdot \nabla \theta \right)_{\theta_0} \left. \frac{\mathrm{d}}{\mathrm{d}\theta}  \right|_{\theta_0} \frac{\omega_D}{v_\| \mathbf{b} \cdot \nabla \theta} \right|  } \right\rangle_{\perp,t}.
   \end{equation}
\editb{The origin of the $v^{-3}$ scaling of \cite{hauff_electrostatic_2009} is now apparent by counting powers of $v$, noting that $J_0\left( x \gg 1 \right) \sim x^{-1/2}$ and $\omega_D \propto v^2$. It was with this scaling that the isotropic slowing-down distribution function was found in Eq. \eqref{Fsdmodef}.}

   \section{\editb{Pitch-angle dependence and anisotropic distribution}} \label{appB}

   \editb{In this section, the pitch-angle dependence of the approximate diffusion coefficient in Eq. \eqref{fluxapprox} is kept to obtain a slowing-down distribution that includes pitch-angle dependent transport. Thus an anisotropic equilibrium distribution for weakly collisional energetic particles is derived. Doing so will require making further assumptions about the turbulent spectrum that should have minimal impact at high energy. }
%   Using the pitch-angle dependent diffusion coefficient of Appendix \ref{appA}, one is equipped to generalize the isotropic distribution treated in the main body of the text. 
   By using the value of the diffusion coefficient at $\xi = 0$, it will be found that Eq. \eqref{Fsdmodef} remains an adequate approximation to the anisotropic distribution up to $|\xi| \approx 0.9$. 

\begin{figure}
      \begin{center}\includegraphics[width=0.5\textwidth]{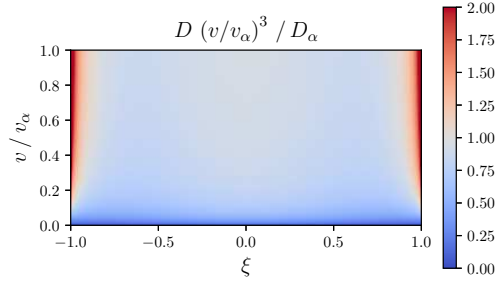} \end{center}
      \caption{\label{Dpa} Velocity dependence of the diffusion coefficient, Eq. \eqref{fluxform}, as compared to the isotropic form, Eq. \eqref{Dmodeldef}. Here, $v_c = 0.3 v_\alpha$ and $v_{ti} =  0.15 k_\mathrm{max} \rho_i v_\alpha$. }
   \end{figure}

   To facilitate calculating the perpendicular spatial average in Eq. \eqref{fluxapprox}, make the ansatz that the turbulent spectrum is on the ion Larmor radius scale and has the following form:
   \begin{equation} \label{phiform}
      \left|\phi\left(\mathbf{k}_\perp \right) \right|^2 = \left| \phi_0 \right|^2_\mathrm{max} \delta\left(k_x \right) \frac{k_y^2}{k_\mathrm{max}^2} e^{- k_y^2 / k_\mathrm{max}^2 + 1}
   \end{equation}
   where $k_y \equiv a^{-1}  \partial S / \partial \alpha$ and $k_\mathrm{max} \sim \rho_i^{-1}$ is the wavenumber at the peak of the turbulent spectrum. The exact form of the spectrum does not significantly impact the functional dependence of the result at high energy (especially when $D_\alpha$ is assumed given as it is here), but provides a meaningful cutoff that will prove necessary. Integrating over all $k_y$ smooths the oscillations in the Bessel function and $\omega_D \propto k_y \left(v_\|^2 + v_\perp^2/2 \right)$. Equation \eqref{fluxapprox} is thereby well-approximated by:
   \begin{equation} \label{fluxform}
      D_{rr} \approx  D_\alpha \frac{v_\alpha^3}{v^3} \frac{1}{1+\xi^2} \frac{1}{\sqrt{1 - \xi^2} + \frac{v_{ti}}{v} \frac{1}{2 \sqrt{\pi} k_\mathrm{max} \rho_i}},
   \end{equation}
   where $D_\alpha$ is the diffusion coefficient at $\xi = 0$ and $v= v_\alpha \gg v_{ti}$. The rightmost factor comes from the Pad\'e approximation $\int_0^\infty k^3 e^{-k^2} J_0^2\left(ak \right) \,\mathrm{d}k \approx \left(4 \sqrt{\pi} a + 2 \right)^{-1}$. This approximate form contains all the velocity dependence of the diffusion coefficient, which is shown in Fig. \ref{Dpa}. Equation \eqref{fluxform} recovers the energy dependence of both electrostatic limits considered in \cite{hauff_electrostatic_2009}, but avoids singularities in pitch angle. The pitch angle dependence in Eq. \eqref{fluxform} is approximately constant except for a narrow region near $\xi = \pm 1$. There, the diffusion coefficient becomes large and the energy dependence changes to $v^{-2}$ \citep{pueschel_anomalous_2012}. This represents the part of phase space where the energetic particle's Larmor radius becomes comparable to $\rho_i$, despite its large energy, by virtue of its large pitch-angle. In this regime, the details of the turbulence do matter and the ansatz made in Eq. \eqref{phiform} becomes unreliable. In any case, one can nevertheless expect a significant increase in the diffusion coefficient when $v \approx |v_\||$. 

\begin{figure}
      \includegraphics[width=1.0\textwidth]{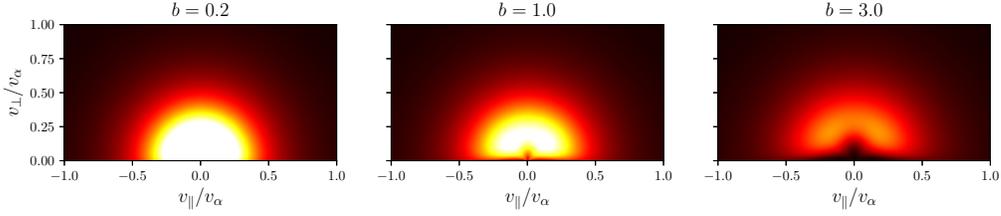}
      \caption{\label{Fb} Pitch angle dependence of the generalized slowing-down distribution, Eq. \eqref{fpadep}, for several values of turbulent intensity (parametrized by $b$). Here, $v_c = 0.3 v_\alpha$ and $v_{ti} = 0.15 k_\mathrm{max} \rho_i v_\alpha$. }
   \end{figure}

\begin{figure}
      \begin{center}\includegraphics[width=0.5\textwidth]{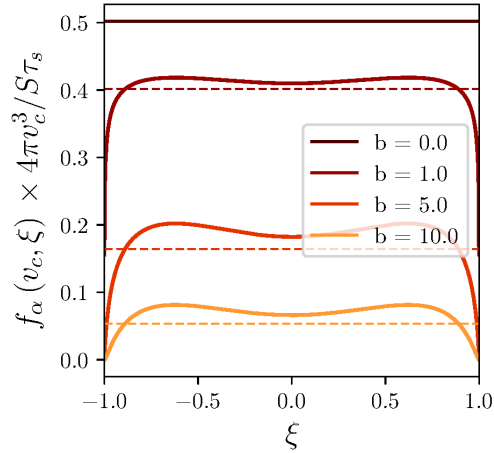} \end{center}
      \caption{\label{padepfig} Pitch angle dependence of the generalized slowing-down distribution, Eq. \eqref{fpadep}, at $v = v_c = 0.3 v_\alpha$ and $v_{ti} = 0.15  k_\mathrm{max} \rho_i v_\alpha$. Dotted lines represent the value of the isotropic distribution, Eq. \eqref{Fsdmodef} using the diffusion coefficient at $\xi = 0$. }
   \end{figure}

   Consider the low-collisionality transport equation with pitch angle scattering included as well as the $\xi$ dependence of transport given by Eq. \eqref{fluxform}:
   \begin{align}    
      b \frac{v_c^3}{v^3}& \left[\left(1 + \xi^2 \right) \left( \sqrt{1-\xi^2} + \frac{v_{ti}}{2 v \sqrt{\pi} k_\mathrm{max}\rho_i} \right) \right]^{-1} f_\alpha \\
      &=  \frac{1}{2} \frac{v_{ti}^3}{v^3} \pd{ }{\xi} \left( 1- \xi^2 \right)  \pd{f_\alpha}{\xi} +  \frac{1}{v^2} \pd{ }{v} \left[ \left( v_c^3 + v^3 \right) f_\alpha \right] + \frac{S_0 \tau_s}{4 \pi v_\alpha^2} \delta\left( v- v_\alpha \right)  \nonumber,
   \end{align}
   where the $v \gg v_{ti}$ limit was used for the deflection collision frequency on the Lorentz operator. Note that, unless $\partial \ln f_\alpha / \partial \xi \gtrsim \mathcal{O}\left(v_c^3 / v_{ti}^3 \right)$, the pitch-angle scattering term is small. In this case, a complicated, but analytic $f_\alpha$ can be found with the help of the integrating factor $\left(v_c^3 + v^3\right)\exp\left(g \right)$, where

   \begin{align} 
      g\left(v, \xi \right) =&  \frac{- b }{\left(1+\xi^2 \right)} \frac{\pi}{24 \pi^2 v_c^3 \left(1-\xi\right)^{3/2} - 3\sqrt{\pi} v_{ti}^3} \times \label{gdef} \\
      & \left\{ 2 \sqrt{3} v_c v_{ti} \left(2 \sqrt{\pi} v_c \sqrt{1 - \xi^2} - v_{ti} \right) \tan^{-1}\left( \frac{2v - v_c}{\sqrt{3} v_c} \right) \right.  \nonumber \\ 
      & + v_c v_{ti} \left(  2 \sqrt{\pi} v_c \sqrt{1 - \xi^2} +  v_{ti}  \right) \ln \left[ \frac{v^2 - v v_c + v_c^2}{\left(v_c + v \right)^2} \right] \nonumber \\
      & \left. -  8 \pi v_c^3 \left(1 -\xi^2\right) \ln\left[ \frac{ v_c^3 + v^3 }{\left(2 \sqrt{\pi} v \sqrt{1 - \xi^2} + v_{ti}   \right)^3} \right] \right\}. \nonumber 
   \end{align}
   The approximate distribution function with the full pitch-angle dependence of Eq. \eqref{fluxform} is therefore:
   \begin{equation} \label{fpadep}
      F_{SD,\mathrm{mod}}\left(v, \xi \right) = \frac{S_0 \tau_s}{4 \pi} \frac{1}{v_c^3 +v^3 } \exp\left[g\left( v_\alpha, \xi \right) - g\left( v, \xi \right) \right] H\left( v_\alpha - v \right) .
   \end{equation}
   This distribution is shown for select parameters in Fig. \ref{Fb}, and the pitch angle dependence is shown in Fig. \ref{padepfig}. With this form, the assumption of negligible pitch angle scattering breaks down in the region $|\xi| \gtrsim 0.99$, where pitch angle derivatives become sufficiently large compared to the distribution. Also in this region, the turbulent spectrum used to derive Eq. \eqref{fpadep} becomes unreliable because the transport will depend on the details of the turbulence, which can only be found from direct simulation.

\iffalse
   To get an idea for the behavior of the distribution near $\xi = \pm 1$, consider the extreme case where $\left( 1- \xi^2 \right) \ll v_{ti}^2 / v^2$ so that the pitch angle dependence in Eq. \eqref{fluxform} is suppressed. Then, Eq. \eqref{patransport} becomes a Legendre equation in this narrow boundary layer, whose solution is
   \begin{equation} \label{blsoln}
      f^{(\RN{3})}_\alpha \approx c^{(\RN{3})}(v) P_L\left( \xi \right),
   \end{equation}
   where $P_L$ is the Legendre function of the first kind and
   \begin{equation} \label{Ldef}
      L \left( L + 1 \right) \approx L^2 = \frac{\sqrt{\pi}}{2} b \frac{v}{v_{ti}} \frac{v_c^3}{v_{ti}^3} k_\mathrm{max} \rho_i \gg 1.
   \end{equation}
   The solution in the intermediate region, $f^{(\RN{2})}$, where both pitch angle scattering and the $\xi$ dependence of Eq. \eqref{patransport} are important, would be asymptotically matched to Eqs. \eqref{fpadep} and \eqref{blsoln}. However, Eq. \eqref{blsoln} is sensitive to the spectrum of the turbulence.
\fi

\section{\editc{Global model}} \label{globalapp}

In this section, an analytic distribution with radial dependence is derived similarly, but more generally, than \cite{anderson_influence_1991}, which will aid in interpreting the length scale $L_\alpha$ more rigorously. This is possible for a fictitious radial profile in the large aspect ratio limit where $D_{rr}$, $\tau_s$, and $v_c$ are independent of radius, while the energetic particle source (which is typically more strongly peaked than the other parameters) retains full radial dependence. This establishes that the velocity dependence of the model maintains its general form even when consistently accounting for radial transport. Equation \eqref{fullsoln} provides a turbulence-modified analytic distribution with radial, energy, and pitch-angle dependence.

Expand the radial dependence of the source in a Fourier-Bessel series:
\begin{equation}\label{sourcexp}
   S_\alpha\left(r, v\right) =  \sum\limits_{j=1}^\infty \frac{\sigma_j}{4 \pi v_\alpha^2} J_0\left( a_{0,j}\frac{r}{a} \right)
\end{equation}
where $\sigma_1$ is the velocity-integrated source at the magnetic axis and $a_{0,j}$ is the $j$th positive root of the zero-order Bessel function $J_0$. This diagonalizes the radial diffusion operator in the large-aspect ratio limit where $V'\left(r \right) \propto r$. Expand $f_\alpha\left(r,v\right) = \sum f_{\alpha,j}\left(v\right) J_0\left(a_{0,j} r /a \right)$ and use the same approximations of Section \ref{section2} for radial diffusion, the collision operator, and source to obtain:
\begin{align} 
-\sum\limits_{j=1}^\infty  & f_{\alpha,j} \frac{1}{r} \pd{}{r}  D_\alpha \frac{v_\alpha^3}{v^3} \pd{}{r} J_0 \left( a_{0,j} \frac{r}{a} \right)  + \frac{1}{\tau_s v^2} \pd{ }{v} \left[ \left( v_c^3 + v^3 \right) f_{\alpha,j} \right] J_0 \left( a_{0,j} \frac{r}{a} \right) \\
   &= \sum\limits_{j=1}^\infty \frac{\sigma_j}{4 \pi v_\alpha^2} \delta\left(v - v_\alpha \right)  J_0 \left( a_{0,j} \frac{r}{a} \right) = 0. \label{besselr} 
\end{align}
After multiplying by $rJ_0\left(a_{0,k} r / a\right)$ and integrating from $r=0$ to $a$, the approximate equation for each Bessel mode $k$ is:
\begin{equation} \label{transpexp}
   D_\alpha \frac{a_{0,k}^2}{a^2} \frac{v_\alpha^3}{v^3} f_{\alpha, k} - \frac{1}{\tau_s v^2} \pd{ }{v}\left[ \left( v_c^3 + v^3 \right) f_{\alpha,k} \right] + \frac{\sigma_k}{4 \pi v_\alpha^2} \delta\left(v - v_\alpha \right).
\end{equation}
This can be solved the same way as Eq. \eqref{approxtransporteqn} to obtain:
\begin{equation}\label{radialsoln}
   f_{SD,r} \left( r, v \right) = \sum\limits_j \frac{\sigma_j \tau_s}{4 \pi} \frac{1}{v_c^3 +v^3 } \left( \frac{v^3}{v_\alpha^3} \frac{v_\alpha^3 + v_c^3}{v^3 + v_c^3} \right)^{b_j/3 }  H\left( v_\alpha - v \right) J_0\left( a_{0,j} \frac{r}{a} \right)
\end{equation}
where:
\begin{equation}\label{bjdef}
   b_j \equiv \frac{D_\alpha \tau_s}{a^2} \frac{v_\alpha^3}{v_c^3} a_{0,j}^2,
\end{equation}
and
\begin{equation}\label{sigmadef}
   \sigma_j \equiv \frac{2\int_0^a S_0(r) J_0\left(a_{0,j} r / a \right)r\, \mathrm{d}r}{a^2 J_1^2\left(a_{0,j}\right) }.
\end{equation}
This is a radially global distribution approximately valid where the plasma profile (particularly, $D_{rr}$, $\tau_s$, and $v_c$) is uniform, but the energetic particle source is not. It has a similar velocity dependence to the local distribution, Eq. \eqref{Fsdmodef}. If only the first term is used to approximate the source, then $S_\alpha \propto J_0\left( a_{0,1} r /a \right)$, and the correspondence is direct with a more rigorous definition of $L_\alpha = a / a_{0,1}$. Otherwise, the global distribution will be a sum of distributions like  $F_{SD,\mathrm{mod}}$, with each term having its own value of $b$.

If the pitch-angle and radial dependence are retained simultaneously and one further assumes that $v_{ti}$ is uniform in $r$, an analytic distribution is obtainable in 3D phase space:
\begin{equation}\label{fullsoln}
   f_{SD,3D} \left( r, v, \xi \right) = \sum\limits_j \frac{\sigma_j \tau_s}{4 \pi} \frac{1}{v_c^3 +v^3 } \exp\left[ g_j\left(v_\alpha, \xi \right) - g_j \left(v, \xi \right) \right]  H\left( v_\alpha - v \right) J_0\left( a_{j,0} \frac{r}{a} \right), 
\end{equation}
where $g_j$ is given by Eq. \eqref{gdef}, except $b$ replaced by $b_j$ from Eq. \eqref{bjdef}.

\bibliographystyle{jpp}
\bibliography{zotero}

\end{document}